# Input Devices for Musical Expression: Borrowing Tools from HCI

**Nicola Orio**
Real-Time Systems Group
Ircam — Centre Pompidou
1, Place Igor Stravinsky
75004 — Paris - France
Nicola.Orio@ircam.fr

**Norbert Schnell**
Real-Time Systems Group
Ircam — Centre Pompidou
1, Place Igor Stravinsky
75004 — Paris - France
Norbert.Schnell@ircam.fr

**Marcelo M. Wanderley**
Analysis-Synthesis Team
Ircam — Centre Pompidou
1, Place Igor Stravinsky
75004 — Paris - France
mwanderley@acm.org

**ABSTRACT**

This paper reviews the existing literature on input device evaluation and design in human-computer interaction (HCI) and discusses possible applications of this knowledge to the design and evaluation of new interfaces for musical expression. Specifically, a set of musical tasks is suggested to allow the evaluation of different existing controllers.

## INTRODUCTION

A substantial amount of material has been published in the human-computer interaction (HCI) literature on the evaluation of existing input devices as well as on the design of new ones. This material includes works on the definition of representative tasks to be used in the comparison of different devices [2], the use of analytical models of aimed movement [12] [7] [8] [1], and the suggestion of various taxonomies of input devices [2][5].

Live performance of computer music can be seen as a highly specialized field of HCI, dealing with such specific topics as *simultaneous multiparametric control*, *timing* and *rytm*, and *training*. Compared to the commonly accepted approach to the design of input devices in HCI, the design of *input devices for musical expression* (here referred to as *controllers*) has traditionally been marked by an idiosyncratic approach. Although various controllers have been proposed [14] [13] they have been usually developed in response to precise artistic demands.

The design of controllers for interactive systems have benefit from an unusually high amount of creativity, in particular if compared to better structured fields where the tendency to follow guidelines may inhibit the appearance of innovative designs [3].

The counterpart of this creativity is the lack of commonly accepted methodologies for the evaluation of existing developments, which prevents from the comparison of different controllers and from the evaluation of their performances in different musical contexts.

Results from classical HCI may be used as tools for developing methodologies for the evaluation of controllers, providing that one is aware of substantial differences, such as: the main channel of communication, visual and auditory; the goal of the interaction, a work to be done and artistic expression; the number of potential users and their expected level of expertise.

## EXISTING RESEARCH IN HCI

The problem arising in the evaluation of an input device is the large number of the parameters involved. To overcome this problem, it has been proposed to compare the devices through the analysis of their performances over a set of representative, and simple, tasks. Another approach was to consider, for each task, which is the most suitable device depending either on its mechanical characteristics or on its matching to the perceptual structure of the task.

### Evaluation Tasks and Methodologies

Buxton [2] has proposed the following tasks as a means to evaluate the match of input devices to applications: *pursuit tracking*, *target acquisition*, *freehand inking*, *tracing and digitizing*, *constrained linear motion*, *constrained circular motion*.

Each of the tasks consists of a common user action in HCI, with its own demands. The choice is clearly driven by the application domain, which is the development of graphical user interfaces, and may not be satisfactory in the musical domain. The creation of any kind of task implies the problem of quantification of input device performances in each task. Indeed, the existence of an evaluation methodology for the *target acquisition* made it the most widely used.

### Fitts' law

Fitts proposed a formal relationship to describe human performance - *speed/accuracy tradeoff* - in aimed movements. Equation (1) shows one formulation of the *Fitts' law* [12]:

$$T = a + b \log_2(A/W + 1) \quad (1)$$

Fitts' law predicts that the time needed to point to a target of width *W* at a distance *A* is *T* seconds. Constants *a* and *b* are empirically determined.

Experiments extending the Fitts' model to 2 dimensional tasks have been reported [12], while it has been proposed the use of Fitts' law in the case of navigation, considered as a multiscale pointing [7].

*Fitts' law - applications in HCI*

The main interest of Fitts' law is that it allows the translation of the performance scores from different devices into *indexes of performance*, which are independent from the experimental conditions used in the different tests, allowing a direct comparison of the devices.

The first application of Fitts' law in HCI [4] was the comparison of a mouse, an isometric joystick, and keys in a text selection task. This study has become the reference in this area, influencing subsequent researches. Although Fitts' law is widely used in HCI, there are still various discussions on the meaning of the results obtained [8].

*Meyer's law*

Meyer et al. [15] proposed a relationship describing aimed movements composed of sub-movements:

$$T = a + b\, n\, (A/W)^{1/n} \quad (2)$$

where *n* is the number of sub-movements performed to reach a target of size *W*, at a distance *A* from the hand's initial position, *a* and *b* are constants.

This relationship has been called *Meyer's law*. Fitts' law can be derived from Meyer's law when *n* approaches infinity and it represents the case when subjects can make as many sub-movements as wished.

*Steering law*

Recently a model describing constrained movement performance was proposed. Accot and Zhai [1] developed a technique for the evaluation of trajectory movement tasks based on constrained motion for different path shapes.

The *steering law* for a generic curved path can represented by the following equation:

$$T_C = a + b \int_C ds/W(s) \quad (3)$$

where $T_C$ is the time to move through a curved path *C*, with variable width *W(s)*; *a* and *b* are constants.

**Selection of Input Devices**

It has been proposed to evaluate input devices depending on their mechanical characteristics and their relationship with the perceptual structure of the task.

*Taxonomies of input devices*

The idea behind the proposition of input device taxonomies is to suggest ways of device comparison according to their basic characteristics, for choosing the devices that best fit a given task.

Buxton proposed a taxonomy of *continuous, manually operated* input devices [2]. The main characteristics analyzed are: a) physical variables being sensed (position, motion, or pressure) and b) number of dimensions sensed for each variable.

Another taxonomy was proposed by Card et al. [5]. This taxonomy shows each independent physical variable being sensed and the axis where the action takes place, instead of the whole device. Moreover, only two basic variables (*position* and *force*) and their derivatives are used. Both basic variables are separated as linear or rotary, therefore the equivalent rotary variables to position and force are *angle* and *torque*.

*Integrality versus separability of input devices*

It has been suggested [10] that the evaluation of existing input devices should be shifted from the analysis of their mechanical structure to the evaluation of their fitness to the perceptual structure of the task to be performed.

Multidimensional objects are characterized by their attributes. Attributes that are perceived as combined are considered *integral*, while those that remain distinct are considered *separable*. User tests showed that devices whose control structure match the perceptual structure of the task will perform better.

**APPLICATIONS TO MUSIC**

Regarding the design of controllers, only few attempts have benefit from HCI results. For instance, Vertegaal proposed the comparison of several input devices in a timbre navigation task [17]. In this study, three devices were used to navigate in a four-dimensional timbre space. Users were asked to reach a given timbre with each one. An evaluation of users' movement time and errors was carried out.

As another application, Figure 1 shows a comparison of several controllers using the taxonomy presented in [5]. Six controllers are compared, with respect to the degrees of freedom, the physical variable sensed, and resolution.

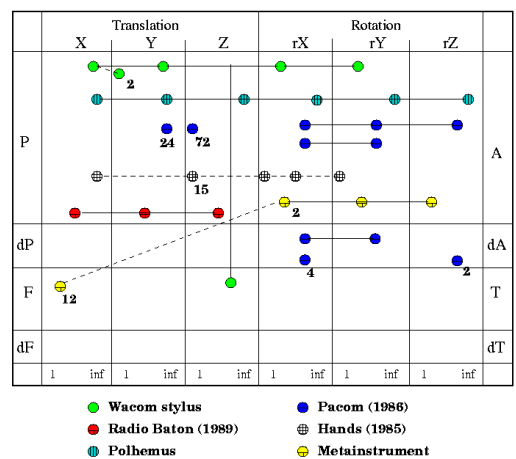

Figure 1: An application of the taxonomy proposed in [5] to various controllers.

**Design Methodologies**

Although most controllers have been designed from an idiosyncratic point of view, a few works have proposed methodologies for the design of musical controllers.

Cariou presented the aXiO, a controller designed from an industrial design perspective [6]. Vertegaal et al. proposed a methodology to match transducer technologies to musical functions, taking into account the types of feedback available with each technology [18]. An evaluation of Vertegaal et al.'s methodology has been presented, where basic musical tasks were defined to validate the proposed relationship [19].

## MUSICAL CONTEXT

The concept of *musical tasks* is part of the process of the evaluation of a musical instrument, since very seldom musicians and composers choose a new instrument without extensively trying it to test how specific musical gestures can be performed. Hence, it looks natural to extend the concept of tasks also to controllers. Research in HCI showed that tasks, in order to be effective, should allow a measurement of performance. In the case of music, it has to be considered that the evaluation of controllers can hardly be done without taking into account subjective impressions of performers, ruled by personal and aesthetic considerations.

From research on HCI, it is apparent that a general feature of musical tasks is simplicity. Even if it may seem totally non-musical, the use of very simple tasks may help in a first step towards the evaluation of controllers. To the statement that music is far beyond the performance of simple musical tasks, it can be replied that writing a novel with a word-processor is far beyond the act of selecting a portion of text.

With the aim of highlighting the most suitable musical tasks, the following considerations are important:

- *Learnability* - It is essential to take into account *the time needed to learn* how to control a performance with a certain controller. It is known that a musician needs more than ten years to master a musical instrument [11], a time far too long for any kind of measurement. Learning to play a second instrument takes less time, because the acquisition of musical ability is not only *kinesthetic*, but also *tonal* and *rhythmic* [16]. Musical tasks thus shall take into account the time needed to learn how to replicate simple musical gestures by experienced musicians.

- *Explorability* - A feature of interest is the *exploration* of the capabilities of the controller, that is, the number of different *gestures* and *gesture nuances* that can be applied. The related musical tasks may require the use of examples the performer is asked to replicate.

- *Feature Controllability* - The accuracy, resolution, and range of features perceived by the user when performing musical tasks. It may happen that a controller will appear totally inadequate for some of the musical tasks, for instance, due to a lack of accuracy.

- *Timing Controllability* - A characteristic of music, which differentiates it from the classical HCI context, is the central role of *time*. This means that musical tasks should also allow measuring temporal precision at which the musician can control the performance and its relationship with the tempo speed.

## SUGGESTED LIST OF MUSICAL TASKS

The most obvious metaphor of interaction in music is the manipulation of an instrument by a performer. Seeing a computer as a musical instrument gives access to a large range of resources of musical literature and traditions for the evaluation of controllers. Even if the metaphor of the musical instrument can be generalized to almost any use of computers in the field of music, many existing applications reproduce a situation that is closer to the interaction between a conductor and an orchestra, which leads to different constraints and observations. Further metaphors can be easily imagined.

Given the previous considerations, a basic, even if incomplete list of musical tasks may be the performance of:

- Isolated tones, from simple triggering to varying characteristics of pitch, loudness, and timbre;
- Basic musical gestures: glissandi, trills, grace notes, and so on;
- Simple scales and arpeggios at different speed, range, and articulation;
- Phrases with different contours, from monotonic to random;
- Continuous feature modulation (e.g. timbre, amplitude or pitch) both for a given note and inside a phrase.
- Simple rhythms at different speeds combining tones or pre-recorded material;
- Synchronization of musical processes.

For each of the above tasks, a measure indicating the degree of polyphony is to be added.

An application of the continuous feature modulation task has been presented in [19], where subjects were supposed to perform a modulation of pitch in each second note of a) a circular path consisting of four notes and b) two pairs of notes placed in different trajectories, as shown below.

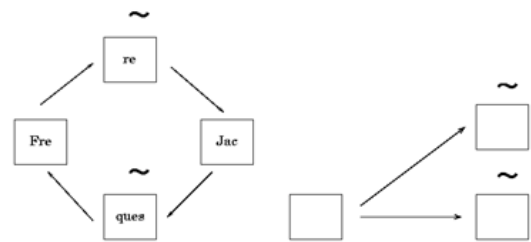

Figure 2: Musical tasks used for the evaluation of the match of transducer technologies and musical functions.

*Comparison with HCI research*

It is easy to draw a parallel between some of the musical tasks and the tasks used in HCI. In particular, target acquisition may be similar to the performance of single tones (acquiring a given pitch as well as a given loudness or timbre [18]), while constrained motion may be similar to the performance of phrase contours. Other musical tasks are peculiar to music, for instance all the ones related to timing and rhythm have no parallel in HCI**.** We believe that in this case it is possible to pinpoint general laws, for instance related to the learning time or the maximum speed allowed by a given controller, that could be useful for future designs. Extensive research based on users may help in the definition of such laws.

The use of musical tasks may also aid the evaluation of existing controllers by defining which is the set of musical gestures that a controller can or cannot perform, together with an indication of the ones each controller perform best.

The definition of a "chart of controllers" that summarizes the main characteristics of available controllers, can be a step towards a more systematic approach of controllers design and use in music. Nevertheless, we believe that the use of well-defined musical tasks is more suitable for musical aims. This mainly because of the crucial roles of mapping [9] and sound synthesis in the overall performances of a controller, that cannot be analyzed for its mechanical characteristics. The evaluation of a controller as a whole can be done only assuming the user's point of view, that is the one of the musician who is asked to play.

**CONCLUSIONS**

In this paper we have presented a review of various methodologies to evaluate input devices from HCI and discussed their applications to the musical domain. A set of musical tasks for the evaluation of controllers was proposed as an initial step towards a systematization of the field.

We consider that a bi-directional flow of knowledge between classical HCI research on input devices, dealing mostly with pointing and dragging material on graphical interfaces, and the design of new computer-based musical instruments can lead to improvements in both fields.